\documentclass[twocolumn,preprintnumbers,showpacs,amsmath,amssymb]{revtex4}

\usepackage{graphicx,graphics} % include figure files
\usepackage{bm} % bold math
\usepackage{dcolumn}% Align table columns on decimal point

\bibpunct{(}{)}{;}{x}{,}{,}
\usepackage{amsmath}
\usepackage{amssymb}
\usepackage{graphics}
\usepackage{times}

\usepackage{setspace}

\newcommand{\bc}{\begin{center}}
\newcommand{\ec}{\end{center}}
\newcommand{\be}{\begin{equation}}
\newcommand{\ee}{\end{equation}}
\newcommand{\bea}{\begin{eqnarray}}
\newcommand{\eea}{\end{eqnarray}}
\newcommand{\bes}{\begin{eqnarray*}}
\newcommand{\ees}{\end{eqnarray*}}
\newcommand{\bee}{\begin{enumerate}}
\newcommand{\eee}{\end{enumerate}}
\newcommand{\bei}{\begin{itemize}}
\newcommand{\eei}{\end{itemize}}

\newcommand{\mH}{\mathcal{H}}
\newcommand{\mR}{\mathcal{R}}
\newcommand{\mS}{\mathcal{S}}

\renewcommand{\S}{{\rm S}}
\newcommand{\bX}{{\bf X}}

\newcommand{\bE}{{\mathbf E}}

\newcommand{{\bxi}}{{\mbox{\boldmath $\xi$}} }
\newcommand{{\mbt}}{{\mbox{\boldmath $\theta$}} }

\newcommand{{\mbz}}{{\mbox{\boldmath $\vartheta$}} }

\newcommand{{\wmbt}}{\widehat{\mbt}}

\newcommand{\ip}[2]{\left <{#1},{#2}\right >}

\newcommand{\supl}{\sup_{\mbt \in \Theta}}

\def\b#1{\mbox{\boldmath $#1$}}

\begin{document}

%\preprint{[Preprint \# appears here.]}

\title{A New Technique for Finding Needles in Haystacks: \\
A Geometric Approach to Distinguishing Between a New Source and 
Random Fluctuations}

\author{Ramani S. Pilla$^{1}$, Catherine Loader$^{1}$, and Cyrus
Taylor$^{2}$}
%\email{pilla@case.edu (corresponding author)}
%\homepage{stat.case.edu/~pillar}
\affiliation{$^{1}$Department of Statistics $^{2}$Department of
Physics \\ Case Western Reserve University, 10900 Euclid Ave.,
Cleveland, Ohio 44106, USA}
\begin{abstract}
%abstract may contain a maximum of 600 characters including spaces!
We propose a new test statistic based on a score process for
determining the statistical significance of a putative signal that may
be a small perturbation to a noisy experimental background. We derive
the reference distribution for this score test statistic; it has an
elegant geometrical interpretation as well as broad applicability. We
illustrate the technique in the context of a model problem from
high-energy particle physics. Monte Carlo experimental results confirm
that the score test results in a significantly improved rate of 
signal detection. \vspace{-0.1in} 
\end{abstract}
%\keywords{}
\vspace{-0.2in}
\maketitle
One of the fundamental problems in the analysis of experimental data
is determining the statistical significance of a putative signal. Such
a problem can be cast in terms of classical ``hypothesis testing'',
where a null hypothesis \(\mH_0\) describes the background and an 
alternative hypothesis \(\mH_1\) characterizes the signal together 
with the background. A test statistic (a function of the data)
is used to decide whether to reject \(\mH_0\) and conclude that a
signal is present.

The hypothesis test concludes that a signal is present whenever the
test statistic falls in a critical region \(W\). One is interested
in the probability that a signal is found under two scenarios. First,
when the null hypothesis \(\mH_0\) is true, the {\em significance level}
\(\alpha\) is the probability of incorrectly concluding that a
signal is present. Second, when the alternative \(\mH_1\) is true, the
{\em power} of the test is the probability that the signal is found.
The goal is to construct a test statistic whose asymptotic
distribution (reference distribution under $\mH_0$ for large sample
size) can be calibrated accurately and that the associated test has
high power at a fixed significance level, such as \(\alpha = 0.01\).

When the two hypotheses are distinct, a powerful technique based on
the likelihood ratio test (LRT) is often used.  Suppose $p(x; \mbt)$
is a probability density function for a measurement $x$ with a
parameter vector $\mbt \in \Theta \subset \mR^d$.  The joint
probability density function evaluated with $n$ measurements $\bX$ for
an unknown $\mbt$ is the likelihood function [\ref{wilks:44}]
$L(\mbt|\bX)$. An effective approach to the problem of choosing
between $\mH_0$ [corresponding likelihood $L(\mbt_0|\bX)$] and $\mH_1$
[with a likelihood $L(\mbt_1|\bX)$] for explaining the data is to
consider the LRT statistic: $\Lambda =
L(\widehat{\mbt}_0|\bX)/L(\widehat{\mbt}_1|\bX)$, where
$\widehat{\mbt}$ is the value of $\mbt$ that maximizes $L(\mbt|\bX)$ 
[\ref{wilks:44}--\ref{cran:03}]. To employ the LRT, the parsimonious
model under $\mH_0$ (with $s_0$ parameters) must be nested within the
more complicated alternative model under $\mH_1$ (with $s_1$
parameters). For simple models, under regularity conditions, $2 \, 
\log(\Lambda)$ is distributed as the $\chi^2$ distribution with $(s_1
- s_0)$ degrees of freedom under $\mH_0$ [\ref{wilks:44}].

When the alternative hypothesis corresponds to a signal which is a
perturbation of the background, regularity conditions required for
this asymptotic theory are violated, since (a) some of the parameters
under $\mH_0$ are on the boundaries of their region of support and (b)
different parameter values give rise to the same null model. As a
result, the LRT has lacked an analytically tractable reference
distribution required to calibrate a test statistic. Such a difficulty
occurs in many practical applications, for example, when testing for a
new particle resonance of unknown production cross section as the
signal strength must be nonnegative. Hence, the LRT must be employed
cautiously; however, it has been employed in several problems of
practical importance where certain required regularity conditions are
violated [\ref{eadie:71}].  An inappropriate application of the LRT
statistics can lead to incorrect scientific conclusions
[\ref{freeman:99},\ref{prot:02}].

In light of the above difficulties with the LRT, a $\chi^2$
goodness-of-fit test is commonly employed. However, it typically has
less power than might be hoped for as it does not take into account
information about the anticipated form of the signal.  We propose a
new test statistic based on a {\em score process} to detect the
presence of a signal and present its reference distribution. This
score statistic is closely related to the LRT for sufficiently large
sample size.

Consider the model
\bes
  \label{eq:perturb}
  p(x; \eta, \mbt) = (1 - \eta) \, f(x) + \eta \, \psi(x; \mbt),
\ees
where $f(x)$ is a specified {\em null density} and $\psi(x, \mbt)$ is
a {\em perturbation density}. The parameter vector $\mbt$ is the
``location'' of the perturbation, and $\eta \in [0, 1]$ measures the
``strength of the perturbation''. The null hypothesis of no signal
($\mH_0\!: \eta = 0$) implies that $p(x; 0, \mbt) = f(x)$ for all $x$
independently of $\mbt$; hence we are in the scenario (b). In
searching for a new particle resonance, for example, one measures the
frequency of events as a function of energy $E$, modeling it by $p(E;
\eta, E_0)$, where $f(E)$ characterizes the background density and
$\psi(E; E_0) = [\Gamma/(2 \, \pi)] [(E - E_0)^2 + (\Gamma/2)^2
]^{-1}$ is the Cauchy (Breit-Wigner) density describing a resonance
centered on $E_0$ with full width at half-maximum $\Gamma$. In this
scenario, $\eta = 0$ under $\mH_0$ and hence the asymptotic
distribution of $2 \, \log(\Lambda)$ under $\mH_0$ does not have an
asymptotic $\chi^2$ distribution.  The asymptotic reference
distribution is not analytically tractable, and hence it is not
possible to employ its measured value for valid statistical inference.

\begin{figure}[htp]
   \centerline{\scalebox{0.6}{\includegraphics{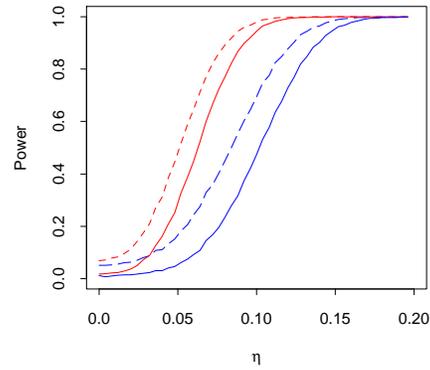}}}
   \caption{(color) Power comparison of the $\chi^2$ goodness-of-fit
   test (blue) and normalized score test ${\mathbb T}$ (red) for $d =
   2$ at $\alpha = 0.05$ (dashed) and $\alpha = 0.01$ (solid),
   calculated {\em via} the volume-of-tube formula, based on 10,000
   simulations for binned data.} \label{fig:power}
\end{figure}

A key obstacle to detecting the signal is finding the tail
probability. We provide an asymptotic solution to this problem {\em
via} a geometric formula (see Eq.\ [\ref{eq:oned}]). The relative
improvement of the score test over the $\chi^2$ goodness-of-fit test
is particularly salient when the signal is hard to detect (see Fig.\ 
\ref{fig:power}).  The development of the reference distribution and a
flexible computational method will enable making probabilistic
statements to solving some of the fundamental problems arising in many
experimental physics.

Pilla and Loader [\ref{pilla:03}] have developed a general theory and
a computationally flexible method to determine the asymptotic
reference distribution of a test statistic under $\mH_0$. Their method
is based on the ``score process'', indexed by the parameter vector
\(\mbt\) and defined as $\S(\mbt) := \partial \, \log[\prod_{i = 1}^n
p(E_i; \eta, \mbt)]/\partial \eta \big|_{\eta = 0}$ for a given data
$\bE = (E_1, \ldots, E_n)$. Under \(\mH_0\), the expectation of
$\S(\mbt)$ is \(0\) for all \(\mbt\), while under $\mH_1$ it has a
peak at the true value of \(\mbt\). Hence, the statistic $\S(\mbt)$ is
sensitive to the signal of interest. The random variability of
$\S(\mbt)$ can exhibit significant dependence on the parameter vector
$\mbt$, hence we consider the {\em normalized score process} defined as
\vspace{-0.1in}
\bea
  \label{eq:nscore}
  \S^{\star}(\mbt) &:=& \frac{\S(\mbt)}{\sqrt{ n \, C(\mbt,
  \mbt)}},
\eea
where \(n\) is the total number of events observed, and
\vspace{-0.1in}
\bea
  \label{eq:covfn}
  C(\mbt, \mbt^{\dag}) &=& \int \frac{
    \psi(x; \mbt) \, \psi(x; \mbt^{\dag})}{f(x)} \, dx - 1
\eea
is the covariance function of \(\S(\mbt)\) for $\mbt \in \Theta
\subset \mR^d$.

For exposition, we assume that $f(E)$, the density under $\mH_0$,
is completely specified. In practice, it often contains unknown
parameters. In this scenario, the covariance function $C(\mbt,
\mbt^{\dag})$ in Eq.\ [\ref{eq:covfn}] for $\S(\mbt)$ needs
modification.  Pilla \& Loader [\ref{pilla:03}] derive an appropriate
$C(\mbt, \mbt^{\dag})$ under estimated parameters.

For testing the hypotheses $\mH_0\!: \eta = 0$ (no signal) versus
$\mH_1\!: \eta > 0$ (signal is present) consider the test statistic
$\mathbb{T} := \sup_{\mbt} \, \S^{\star}(\mbt)$ for $\mbt
\in \Theta \subset \mR^d$. It is concluded that a signal is present if
\(\mathbb{T}\) exceeds a critical level \(c \in \mR\).  The problem
now is to determine the reference distribution of \(\mathbb{T}\), so
that \(c\) can be chosen to achieve a specified significance
level $\alpha$.

Under $\mH_0$, \(\S^{\star}(\mbt)\) converges in distribution to a
Gaussian process \(Z(\mbt)\) with mean 0 and covariance function
$C(\mbt, \mbt^{\dag})/\sqrt{C(\mbt, \mbt) C(\mbt^{\dag},
\mbt^{\dag})}$ as $n \rightarrow \infty$ [\ref{pilla:03}]. The
reference distribution of \(\mathbb{T}\) converges to that of
$\sup_{\mbt} \, Z(\mbt)$ as $n \rightarrow \infty$ for $\mbt \in
\Theta \subset \mR^d$.  
\begin{figure}[htb]
   \centerline{\scalebox{0.85}{\includegraphics{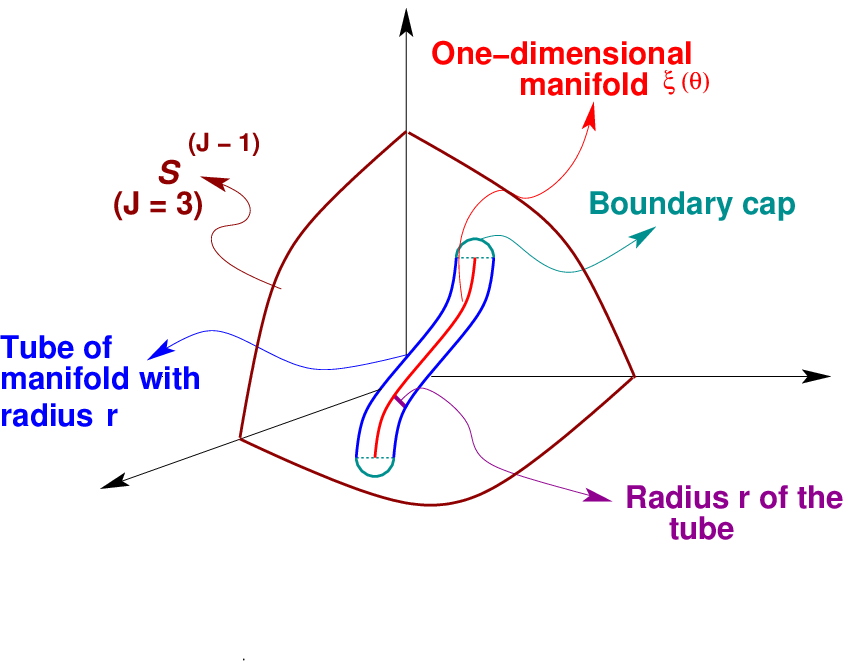}}}
   \vspace{-0.2in}
   \caption{(color) Tube around a one-dimensional manifold
   $\bxi(\mbt)$, with boundaries, embedded in $\mS^2 \subset \mR^3$.}
   \label{fig:tube}
\end{figure}
Except in special cases, this distribution cannot be expressed
analytically. However, a good asymptotic solution to the tail
probability $P(\sup_{\mbt} \, Z(\mbt) \ge c)$, where $c
\in \mR$ is large, can be obtained {\em via} the {\em volume-of-tube}
formula [\ref{hot:39}--\ref{naiman:90}]. The volume-of-tube formula
provides an elegant geometric approach for solving problems in
simultaneous inference [\ref{know:89}] by reducing the evaluation of
tail probabilities to that of finding the $(J - 1)$-dimensional volume
of the set of points lying within a distance $r$ of the curve $(d =
1)$ or {\em manifold} $(d \geq 2)$ on the surface of the unit sphere
in $J$-dimensions for some integer $J$ (see Fig.\ \ref{fig:tube}).

Suppose $\bxi(\mbt)$ defines a manifold for $\mbt$ on the surface of
a $(J - 1)$-dimensional unit sphere $\mS^{(J - 1)}$. Fig.\
\ref{fig:tube} shows a ``tube'' of radius $r$ around a manifold
$\bxi(\mbt)$ embedded in $\mS^{(J - 1)} \subset \mR^J$ with boundary
caps. We represent the Gaussian random field $Z(\mbt)$, {\em via} the 
Karhunen-Lo\`{e}ve expansion [\ref{adler:90}] as $Z(\mbt) =
\sum_{k = 1}^{\infty} \vartheta_k \; \xi_k(\mbt) =
\ip{\mbz}{\bxi(\mbt)}$, where $\ip{\cdot}{\cdot}$ denotes the inner
product, $\mbz$ and $\bxi$ are vectors and $\vartheta_k \sim N(0,
1)$. If the Karhunen-Lo\`{e}ve expansion is terminated after $J$
terms, then the following relation between the manifold $\bxi(\mbt)$
embedded in $\mS^{(J - 1)} \subset \mR^J$ and the Gaussian random
field $Z(\mbt)$ holds [\ref{pilla:03}]:
\vspace{-0.1in}
\begin{eqnarray*}
  \lefteqn{P\left(\underset{\mbt \in \Theta}{\sup} \, Z(\mbt) \ge
  c \right)} \\
   &=& \int_{c^2}^{\infty} \; P\left(\underset{\mbt \in
  \Theta}{\sup} \, \ip{\b U}{\bxi(\mbt)} \ge \;  w \right)
  h_J(y) \; dy,
\end{eqnarray*}
where $\b U = (U_1 = \vartheta_1/\|\mbz\|, \ldots, U_J =
\vartheta_J/\|\mbz\|)$ is uniformly distributed on $\mS^{(J - 1)}
\subset \mR^{J}$, $\bxi = (\xi_1, \ldots, \xi_J)$, $w =
c/\sqrt{y}$, and $h_J(y)$ is a $\chi^2$ density with $J$ degrees of
freedom. The uniformity property enables finding the $P(\cdot)$ in the
integrand {\em via} the volume-of-tube formula. Note that $r^2 = 2 (1
- w)$.

Geometrically, $P( \sup_{\mbt} \, \ip{\b U}{\bxi(\mbt)} \ge
\; w)$ is the probability that $\b U$ lies within a tube of radius $r$
around $\bxi(\mbt)$ on the surface of $\mS^{(J - 1)}$ and equals the
volume of tube around $\bxi(\mbt)$ divided by the surface area of
$\mS^{(J - 1)}$ [\ref{hot:39}, \ref{weyl:39}].
In effect, constructing a test of significance level 5\% is
equivalent to choosing the rejection set covering 5\% of $\mS^{(J -
1)}$. Therefore, finding critical values of the test statistic
${\mathbb T}$ is equivalent to finding a $(J - 1)$-dimensional volume
of the tube.

The results of Hotelling-Weyl-Naiman [\ref{hot:39}--\ref{naiman:90}]
imply that for $w \approx 1$, the tail probability is expressible as a
weighted sum of \(\chi^2\) distributions, with $(d + 1)$ terms and
coefficients that depend on the geometry of the $d$-dimensional
manifold $\bxi(\mbt)$. The results of Pilla and Loader
[\ref{pilla:03}] provide an expansion of the distribution of
$\sup_{\mbt} \, Z(\mbt)$ in terms of the \(\chi^2\) probabilities:
\vspace{-0.1in}
\begin{eqnarray}
  \label{eq:oned}
   \lefteqn{P\left( \supl \, Z(\mbt)  \ge c \right)} \nonumber \\
  &=& \; \sum_{k = 0}^d \frac{\zeta_k}{A_k A_{d + 1 - k}} P\left(
  \chi_{d + 1 - k}^2 \ge c^2 \right) \nonumber\\ 
  && + \; o(c^{-1/2}e^{-c^2/2}) \; \mbox{as} \; c \to\infty,
\end{eqnarray}
where \(A_0 = 1\) and \(A_k = 2 \, \pi^{k/2}/\Gamma(k/2)\) for $k \ge
1$. The constants \(\zeta_0, \ldots, \zeta_d\) depend on the geometry
of the $\xi(\mbt)$; \(\zeta_0\) is the area of the manifold and
\(\zeta_1\) is the length of the boundary of the manifold. These can
be represented explicitly in terms of the covariance function: 
\vspace{-0.1in}
\[
  \hspace{-0.1in}
  \zeta_0 = \int_{\mbt \in \Theta}
    [C(\mbt, \mbt) ]^{-\frac{(d + 1)}{2}} \, D(\mbt, \mbt)  \,
      d \mbt,
\]
where $D(\mbt, \mbt)$ is defined as 
\vspace{-0.1in}
\[
 \Bigg|{\det}
       \begin{pmatrix}
       C(\mbt, \mbt^{\dag}) &
       \nabla_1 \, C(\mbt, \mbt^{\dag}) \\  \nabla_2 \,
     C(\mbt, \mbt^{\dag}) & \nabla_1 \nabla_2 \, C(\mbt,
     \mbt^{\dag})
        \end{pmatrix} \Bigg|_{\mbt^{\dag} = \mbt}^{\frac{1}{2}}
\]
with \(\nabla_1\) and \(\nabla_2\) as the partial derivative operators
with respect to \(\mbt\) and \(\mbt^{\dag}\) respectively. The
expression for \(\zeta_1\) is similar except that integration is over
the boundary of the manifold. The remaining constants involve
curvature of the manifold and its boundaries, and become progressively
more complex. However, for practical problems the first few terms will
suffice and an implementation of the first four terms is described in
[\ref{loader:04}]. When the reference distribution can be approximated
by a $\chi^2$ distribution, then a tabulated value can be employed to
calibrate the test statistic whereas the geometric constants appearing
in the above tail probability evaluation depend on the problem at
hand. In this modern computer era, it is not difficult to compute them
numerically [\ref{loader:04}].

In many applications, including the one considered in this letter, 
one is interested in the probabilities of rare events (i.e., \(c 
\rightarrow \infty\)). In this case, the terms in Eq.\ (\ref{eq:oned})
are of descending size, and the error term is asymptotically
negligible.

\begin{figure}[htb]
  \vspace{-0.1in}
%  \centerline{\scalebox{0.3}{\includegraphics{scoresurf2}}}
\setlength{\unitlength}{1mm}
\begin{picture}(80,80)
\put(40,40){\makebox(0,0){See separate file for Figure 2.}}
\end{picture}
  \caption{(color) Surface of the process $\S^{\star}(\mbt)$ as a
  function of $\mbt = (E_0, \Gamma)$.}
  \label{fig:3d}
\end{figure}
We demonstrate the power of the score test with a Monte Carlo
simulation experiment drawn from high-energy physics. In our
simulation, we consider measurements of energy in a region $E \in [0,
2]$ in which the background (null) density is modeled as linear, with
a specific form $f(E) = (1/2.6) \, (1 + 0.3 E)$. The resonance is
modeled by a Breit-Wigner density function. The parameters for this
problem are modeled following an example in Roe [\ref{roe:92}].

To examine the effectiveness of the test ${\mathbb T}$ in detecting a
signal, we perform Monte Carlo analyses of 10,000 samples each with a
size of $n = 1000$ events spread over 50 bins at the values of $\Gamma
= 0.2$ and $E_0 = 1$. For a single simulated dataset, Fig.\
\ref{fig:3d} shows the normalized score surface as a function of $E_0$
and $\Gamma$. It is clear that the maximum is achieved at $E_0 = 1$ 
irrespective of the value of $\Gamma$.
\begin{figure}[htb]
  \centerline{\scalebox{0.6}{\includegraphics{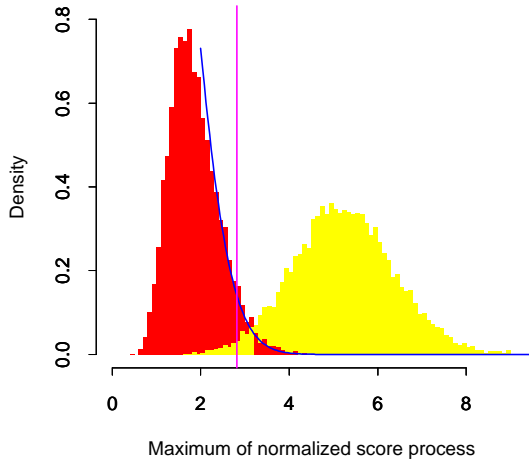}}}
  \caption{(color) Histograms of the simulated null ($\eta = 0$)
  density (red) and alternative ($\eta = 0.1$) density (yellow) of the
  test statistic ${\mathbb T}$ with a superimposed (blue) asymptotic
  null density (derivative of Eq. [\ref{eq:oned}]) for a fixed
  $\Gamma$. The purple vertical bar is the cut off for the test
  statistic ${\mathbb T}$ at the 5\% false positive rate calculated
  {\em via} the volume-of-tube formula (Eq. [\ref{eq:oned}] with $d =
  1$).}  
  \label{fig:hist2}
\end{figure}

Fig.\ \ref{fig:hist2} shows histograms over 10,000 samples under the
$\mH_0\!: \eta = 0$ and $\mH_1\!: \eta = 0.1$ for a fixed
$\Gamma$. The former histogram confirms that about 5\% of the time,
hypothesis of no signal be rejected. The asymptotic null density
(derivative of Eq. [\ref{eq:oned}] with $d = 1$) agrees with the
simulated null distribution as expected.

When both $E_0$ and $\Gamma$ are estimated, Fig.\ \ref{fig:power}
shows that the power of detection increases as the signal
strength $\eta$ increases. Our test statistic ${\mathbb T}$ is
significantly more powerful than the $\chi^2$ goodness-of-fit test in
detecting the signal. The asymptotic tail probability result obtained
{\em via} the volume-of-tube formula (Eq. [\ref{eq:oned}]) is elegant,
simple and powerful in distinguishing the signal and the random
fluctuations in data.

Financial support from the U.S. National Science Foundation, Division
of Mathematical Sciences and the Office of Naval Research, Probability
\& Statistics Program is gratefully acknowledged.

\bee
\vspace{-0.08in}
\item \label{wilks:44}  Wilks, S.S.\ {\em Mathematical
Statistics}. (Princeton University Press, New Jersey, 1944).
\vspace{-0.08in}
\item\label{eadie:71} Eadie W.T.\ {\em et al.}. {\em Statistical
Methods in Experimental Physics} (New York: North-Holland, 1971).
\vspace{-0.08in}
\item \label{cran:03} Cranmer, K.S.\ {\em PHYSTAT2003, SLAC}, Stanford,
California (2003).
\vspace{-0.08in}
\item \label{freeman:99} Freeman, P.E.\ {\em et al.} {\em
Astrophys. J.} {\bf 524}, 1, 753 (1999).
\vspace{-0.08in}
\item \label{prot:02} Protassov, R.\ {\em et al.} {\em
Astrophys. J.} {\bf 571}, 1, 545 (2002).
\vspace{-0.08in}
\item \label{pilla:03} Pilla, R.S.\ \& Loader, C. Technical Report,
Department of Statistics, Case Western Reserve University (2003).
%Preprint at (http://stat.cwru.edu/$\sim$pillar/research) (2003).
\vspace{-0.08in}
\item \label{hot:39} Hotelling, H.\ {\em Amer. J. Math.} {\bf 61},
440 (1939).
\vspace{-0.08in}
\item \label{weyl:39} Weyl, H.\ {\em Amer. J. Math.} {\bf 61}, 461
(1939).
\vspace{-0.08in}
\item \label{naiman:90} Naiman, D.Q.\ {\em Ann. Stat.} {\bf 18},
685 (1990).
\vspace{-0.08in}
\item \label{know:89} Knowles, M.\ \& Siegmund, D.\ {\em
Intl. Stat. Rev.} {\bf 57}, 205 (1989).
\vspace{-0.08in}
\item \label{adler:90} Adler, R.J.\ {\em An introduction to Continuity,
Extrema and Related Topics for General Gaussian Processes}. (Institute
of Mathematical Statistics, Hayward, CA, 1990).
\vspace{-0.08in}
\item \label{loader:04} Loader, C.\ {\em Computing Science and Statistics:
Proc. 36th Symp. Interface} (2004).
\vspace{-0.08in}
\item \label{roe:92} Roe, B.P.\ {\em Probability and Statistics in
Experimental Physics}. (Springer, NY, 1992).\
\eee

\end{document}